\documentclass[11pt]{revtex4}    
\usepackage{amssymb,epsf}
\usepackage{latexsym}
\usepackage{setspace}
\begin{document}

\title{ Expanding $(n+1)$-Dimensional Wormhole Solutions in Brans-Dicke Cosmology}
\author{  E. Ebrahimi
\footnote{email: esmebrahimi@gmail.com} and N. Riazi
\footnote{email: riazi@physics.susc.ac.ir}} \affiliation{Physics
Department and Biruni Observatory, Shiraz University, Shiraz
71454, Iran\\}

\begin{abstract}
We have obtained two classes of $(n+1)$-dimensional wormhole
solutions using a traceless energy-momentum tensor in Brans-Dicke
theory of gravity. The first class contains wormhole solutions in
an open geometry while the second contains wormhole solutions in
both open and closed universes. In addition to wormhole
geometries, naked singularities and maximally symmetric spacetime
also appear among the solutions as special cases. We have also
considered the travesibility of the wormhole solutions and have
shown that they are indeed traverseable. Finally, we have
discussed the energy-momentum tensor which supports this geometry
and have checked for the energy conditions. We have found that
wormhole solutions in the first class of solutions violate weak
energy condition (WEC). In the second class, the wormhole
geometries in a closed universe do violate WEC , but in an open
universe with suitable choice of constants the supporting matter
energy-momentum tensor can satisfy WEC. However, even in this case
the full effective energy-momentum tensor including the scalar
field and the matter energy-momentum tensor still violates the
WEC.

\

pacs: 98.80.-k, 98.80.QC, 04.50.-h
\end{abstract}

 \maketitle
\section{Introduction}
The scalar-tensor theory of gravity was proposed by P. Jordan for
the first time \cite{jordan} in 1950. Since the Einstein's theory
of gravity does not admit Mach's principle, Brans and Dicke tried
to modify Einstein's theory in order to incorporate this
principle. They presented their theory in 1961 which is now named
Brans-Dicke (BD) theory \cite{bd}. BD theory describes gravitation
through a metric tensor $g_{\mu\nu}$ and a massless scalar field
$\phi$. In this theory which involves a dimensionless parameter
$\omega$, the gravitational constant is a function of space-time
coordinates.

Wormholes are objects which connect two distant parts of the same
space-time or even two distinct space-times. Although a wormhole
solution first entered the physics literature in 1916 \cite{reff},
the concept was first considered seriously in 1935 by Einstein and
Rosen \cite{ERB} which was later called Einstein-Rosen bridge. The
word "wormhole" was first time coined by Wheeler \cite{wheeler} in
1957. The main objection against wormholes is that the
energy-momentum tensor which supports these geometries turns out
to be un-natural. This kind of matter is called 'exotic'. Exotic
matter violates the weak energy condition (WEC) and also sometimes
other common energy conditions \cite{bord,haw}. After the  work by
Morris and Thorne \cite{MT}, and subsequent works on traversable
Lorentzian wormholes \cite{MTY} there has been a continued
interest in the wormhole issue. Wormholes have been extensively
investigated from different points of view in the literature
\cite{visser,11,14,15,16,17,18,19,20}.

Recently, it has been discovered that the universe is accelerating
\cite{accelreff}. Following this discovery, there has arisen more
attention to scalar-tensor theories, including BD as a prototype.
The reason is that the scalar degree of freedom  in these theories
can be used to explain some features of dark energy (the cause of
the acceleration). In this paper, we are interested in
investigating some exact wormhole solutions of BD theory in a
cosmological background. Many authors have considered the wormhole
solutions of BD theory. Lorentzian wormholes in Brans-Dicke theory
have been analyzed by Agnese and La Camera \cite{agnes} and Nandi
et. al. \cite{nandi}. Exact rotating wormhole solutions of the BD
theory are rarely found and one can see the work by Matos and
N´u˜nez \cite{mat}. Two classes of massive Lorentzian traversable
wormhole solutions in BD theory were found by He and Kim
\cite{kim}. Some  authors have considered observational aspects of
wormholes. For example, Cramer et. al \cite{cramer} investigated
the possibility of detecting wormholes by use of their
gravitational lensing. Recently Harko et. al.\cite{harko} claimed
that it is possible to detect wormholes via their electromagnetic
radiation spectrum.

 Following the inflation theory by A. Guth \cite{guth}
it has been supposed that the quantum fluctuations in the inflaton
field can be assumed as the seed of large scale structures in the
universe. Then non-trivial topological objects such as microscopic
wormholes may have been formed during that era and then enlarged
to macroscopic objects with expansion of the universe. This
constitutes part of our motivation for studying wormhole solutions
in an evolving cosmological background. To investigate this
problem, we use a modified Robertson-Walker (RW) metric and
consider the solutions generated by a traceless source.

 We organize this paper in the following manner: In Sec.
II, we present the ansatz metric and the resulting solutions in
(n+1)-dimensions. In section III, we consider different features
of the solutions. In section IV, we investigate the corresponding
energy-momentum tensor and determine the exoticity parameter. The
last section is devoted to conclusions and closing remarks.

\section{Action, Field Equations and (n+1)-Dimensional solutions\label{field}}
The action of the Brans-Dicke theory with one scalar field $\Phi$
can be written as
\begin{equation}
I_{G}=-\frac{1}{16\pi}\int_{\mathcal{M}}
d^{n+1}x\sqrt{-g}\left(\Phi {R}-\frac{\omega}{\Phi}(\nabla\Phi)^2
+{\cal L}(m)\right),\label{act1}
\end{equation}
where ${R}$ is the scalar curvature, $\cal L$ is the Lagrangian
density of the matter and $\omega$ is the BD constant. Varying
this action with respect to $g_{\mu\nu}$ and $\phi$, we obtain
\begin{eqnarray}
&&\phi \,G_{\mu\nu}=-8\pi T_{\mu\nu}^{M} - \frac{\omega}{\phi}
\left(\phi_{,\mu}\phi_{,\nu}-\frac{1}{2}g_{\mu\nu}\phi_{,\lambda}\phi^{,\lambda}\right)
-\phi_{;\mu;\nu}+g_{\mu\nu}\Box\phi, \label{eq1}
\\\nonumber & &\Box\phi=\frac{8\pi}{(n-1)\omega+n}T_{\lambda}^{M
\,\lambda}, \label{eq2}
\end{eqnarray}
where $T^M_{\mu\nu}$ is the matter energy-momentum tensor. We are
looking for BD wormhole solutions in a cosmological background. To
this end, we use the metric
\begin{equation}
ds^2= -  dt^2 + R(t)^2\left[(1 +a(r))\,dr^2 + r ^2 d\Omega_{n-1}
^2\right],  \label{metn}
\end{equation}
in which $R(t)$ is the scale factor and $a(r)$ is an unknown
function. One will see that $a(r)$ is related to the shape
function of the wormhole solutions. With this generalization, our
metric is not necessarily homogeneous but it is still isotropic
about the center of the symmetry. It is obvious that this metric
is an extension of the Robertson-Walker (RW) metric and is less
symmetric than it.

In order to start our study, we choose a traceless energy-momentum
tensor. With this assumption and equation (\ref{eq1}) we have

\begin{eqnarray}
&&G_{\mu}^{\mu}= \frac{\omega(n-1)}{2\phi^2}
\phi_{;\mu}\phi^{;\mu},\label{eqtr}
\\\nonumber & &\Box\phi=0.
\end{eqnarray}

Here, our task is to solve these equations simultaneously. To this
end, we use the ansatz $\phi=\phi(r,t)=S(t)P(r)$. It can be shown
that \cite{riazinasr}, in a cosmological background,
$P(r)=constant$ and we are led to $\phi=\phi(t)=S(t)$. In BD
theory $\frac{1}{\phi}$ plays the role of effective gravitational
coupling G. Here, we will try a power law scalar field:

\begin{eqnarray}
\phi(t)=At^d, \label{scafield}
\end{eqnarray}
where $A=constant>0$, because $G$, the Newtonian gravitational
constant is positive. Considering this ansatz, it turns out that
$G_{eff}=\frac{1}{\phi}$, the effective gravitational constant is
just as a function of time.

With these assumptions, we are able to solve the equations
(\ref{eqtr}) and find two different classes of solutions.

Class I solutions:

The first class of solutions reads

\begin{eqnarray}
\phi(t)=At^{\frac{-1+\sqrt{n^2-n(n-1)\omega}}{n\omega-(n+1)}},\label{scafield1}
\end{eqnarray}
\begin{eqnarray}
R(t)=C_{1}t^{\frac{-1+\sqrt{n^2-n(n-1)\omega}}{n(n\omega-(n+1))}+\frac{1}{n}},\label{scalfac1}
\end{eqnarray}
\begin{eqnarray}
1+a(r)=\frac{1}{1+\frac{C_{2}}{r^{n-2}}}.\label{b1}
\end{eqnarray}

In these solutions, $C_{1}$ and $C_{2}$ are integration constants.
These solutions reduce to those presented in \cite{riazinasr} for
$n=3$. It can be easily shown that such a geometry is supported by
the following energy-momentum tensor:

\begin{eqnarray}
\rho=-T_{t}^{Mt}=\alpha_{1}At^{\gamma_{1}} ,\label{rho1}
\end{eqnarray}
\begin{eqnarray}
P_{r}=T_{r}^{Mr}=\alpha_{2}At^{\gamma_{1}}
-\beta_1\frac{C_{2}}{C_{1}^2}\frac{At^{\gamma_2}}{r^n},\label{pr1}
\end{eqnarray}
\begin{eqnarray}
P_{t}=T_{\theta}^{M\theta}=T_{\phi}^{M\phi}=...=\alpha_{2}At^{\gamma_1}+\beta_2\frac{C_{2}}{C_{1}^2}\frac{At^{\gamma_2}}{r^n},\label{pt1}
\end{eqnarray}

where

\begin{eqnarray}
\alpha_{1}=\frac{(n((n-2)\omega-(n-1))+(n\omega-(n-1))\sqrt{n^2-n(n-1)\omega})}{32\pi^2
G(n\omega-(n+1))^2}=n\alpha_2,
\end{eqnarray}
\begin{eqnarray}
\beta_{1}=\frac{(n-1)(n-2)}{128\pi^2 G}=(n-1)\beta_2,
\end{eqnarray}
\begin{eqnarray}
\gamma_{1}=\frac{(2n+1)+\sqrt{n^2-n(n-1)\omega}-2n\omega}{n\omega-(n+1)},
\end{eqnarray}
and
\begin{eqnarray}
\gamma_{2}=\frac{(n+2)\sqrt{n^2-n(n-1)\omega}+n(1-2\omega)}{n(n\omega-(n+1))}.
\end{eqnarray}

Class II of solutions:

The second class of solutions reads

\begin{eqnarray}
\phi(t)=At^{1-n},\label{scafield2}
\end{eqnarray}
\begin{eqnarray}
R(t)=C_{1}t,\label{scalfac2}
\end{eqnarray}
and
\begin{eqnarray}
1+a(r)=\frac{1}{1+\frac{C_{2}}{r^{n-2}}+C_{1}^2\left(1-\frac{(n-1)\omega}{n}\right
)r^2},\label{b2}
\end{eqnarray}
where, once again, $C_{1}$ and $C_{2}$ are integration constants
related to boundary conditions of our problem and determine what
kind of solutions we have. This geometry is supported by the
following energy momentum tensor

\begin{eqnarray}
&&\rho=-T_{t}^{Mt}=-\eta_1 At^{-(n+1)} ,\label{rho2}
\\\nonumber &&P_{r}=T_{r}^{Mr}=-\eta_2At^{-(n+1)}-\kappa_1\frac{C_{2}}{C_{1}^2}\frac{At^{-(n+1)}}{r^n},\label{pr2}
\\\nonumber &&P_{t}=T_{\theta}^{M\theta}=T_{\phi}^{M\phi}=...=-\eta_2At^{-(n+1)}+\kappa_2\frac{C_{2}}{C_{1}^2}\frac{At^{-(n+1)}}{r^n},\label{pt2}
\end{eqnarray}
where
\begin{eqnarray}
\eta_1=\frac{(n-1)(n+(n-1)\omega)}{64\pi^2 G}=n\eta_2,
\end{eqnarray}
and
\begin{eqnarray}
\kappa_1=\frac{(n-1)(n-2)}{128\pi^2 G}=(n-1)\kappa_2.
\end{eqnarray}

\section{Properties of the Solutions\label{field1}}

In order to investigate whether a given solution represents a
wormhole geometry, it is convenient to take a look at the metric
which is used by Morris and Thorne \cite{MT}. In that paper, the
metric of the wormhole is written in the form:

\begin{equation}\label{dr}
    \mathit{\ ds}^{2}= - \mathit{\ d}\,t^{\mathit{2\ }} +
{\displaystyle \frac {\mathit{\ d}\,r^{\mathit{2\ }}}{1 -
{\displaystyle \frac {\mathrm{b}(r)}{r}} }}  +
r^{2}\,\left[\mathit{\ d}\,\theta ^{\mathit{2\ }} +
\mathrm{sin}^{2}(\theta )\,\mathit{ \ d}\,\phi ^{\mathit{2\
}}\right]
\end{equation}
Where $b(r)$ is the shape function and the throat radius satisfies
$b(r_0)=r_0$. If the equation $d(r)=r-b(r)$ has any root $r_0$ and
simultaneously $d(r)>0$ for $r>r_0$ then we will have a wormhole
and $r_0$ gives the throat radius of the wormhole. Such a geometry
can be taken as a wormhole, connecting two distinct universes. We
apply such a procedure to our solutions. Since the ansatz metric
expands with time, the wormhole throat circumference

\begin{equation}
 l =R(t)r_{0}\oint d\phi =2\pi r_{0}R(t)
 \label{throat}
\end{equation}

expands with time in proportion to the scale factor.

Let us investigate the two classes of solutions separately.

\subsection{Class (I)}

According to the equation (\ref{b1}) we have

\begin{equation}
\mathrm{d}(r) =0 \Rightarrow r^{n-2}+C_{2}=0,
 \label{db1}
\end{equation}

where $C_{2}$ is a constant. It is obvious that this equation has
at least one root. In order to study the curvature of this
solution, we calculate Kretschman scalar. It turns out  that this
scalar blows up at the origin ($r=0$). One can see that for
$C_{2}<0$, $d(r)$ has always a root but the radial coordinate $r$
never achieves ($r=0$) and the geometry in this case represents
two open universes, connected by a wormhole. For $C_{2}\geq0$,
$d(r)$ has not any root and the $r$ coordinate can reach the
origin. In this case, therefore, we have a naked singularity in an
open universe.

An interesting point in this class of solutions is that the scale
factor, $R(t)$, and the BD scalar field, $\phi(t)$, depend on the
BD constant $\omega$ while $1+a(r)$ is independent of $\omega$.
This shows that changing the $\omega$, does not affect the
geometry of this class of solutions and only changes the expansion
rate of the universe and the dynamic of the BD scalar field. It is
interesting to note that the $\omega\rightarrow\infty$ limit of
this solution yields $\phi(t)= A$ as we expected, while
$R(t)\propto t^{\frac{1}{n}}$ which is not identical with the
corresponding GR solution.

Let us have a look at the Ricci scalar. The Ricci scalar in
(n+1)-dimensions for class (I) solutions reads

\begin{equation}
{\cal R}=
\frac{\omega(n^2+1-n(n-1)\omega-2\sqrt{n^2-n(n-1)\omega})}{(n\omega-(n+1))^2\,t^2}.
\label{ricci1}
\end{equation}

It can be seen that the Ricci scalar doesn't depend on $r$ and is
only a function of time.

\subsection{Class (II)}
In this subsection, we study its second class of solutions and
discuss the geometrical properties. Once again, the Kretschman
scalar indicates an intrinsic singularity at $r=0$.

\begin{figure}\epsfysize=5cm
{ \epsfbox{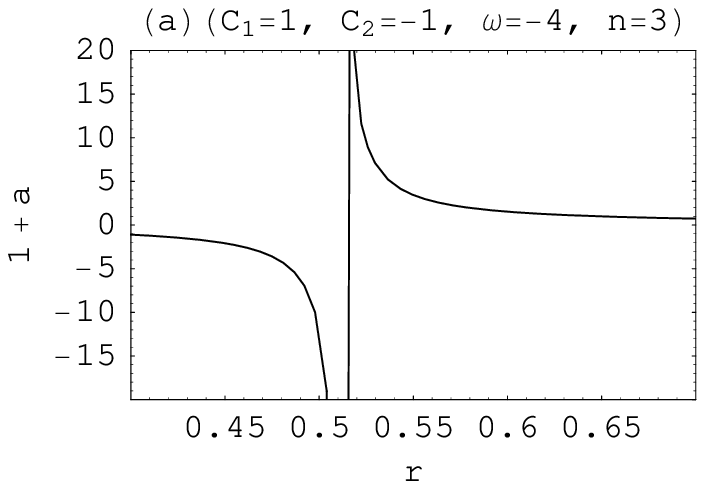}}\epsfysize=5cm { \epsfbox{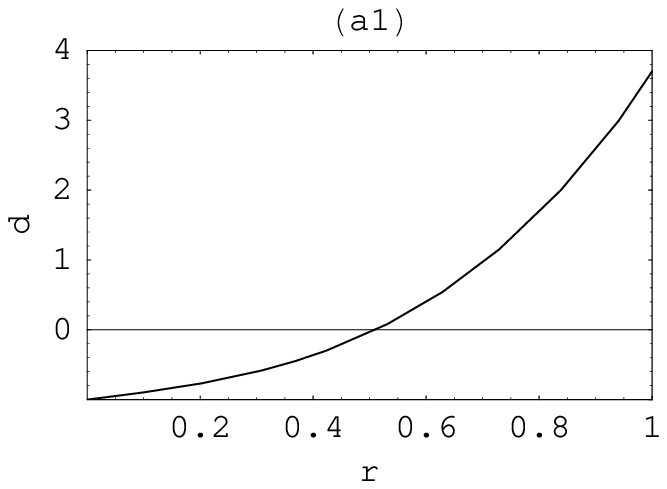}}\epsfysize=5cm
{ \epsfbox{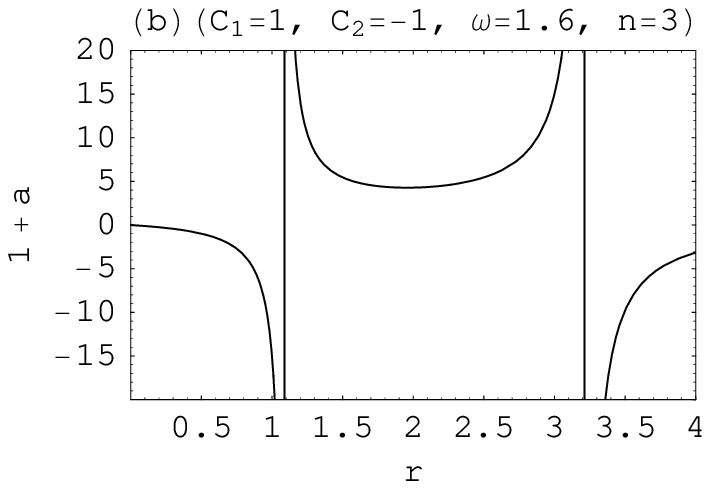}}\epsfysize=5cm { \epsfbox{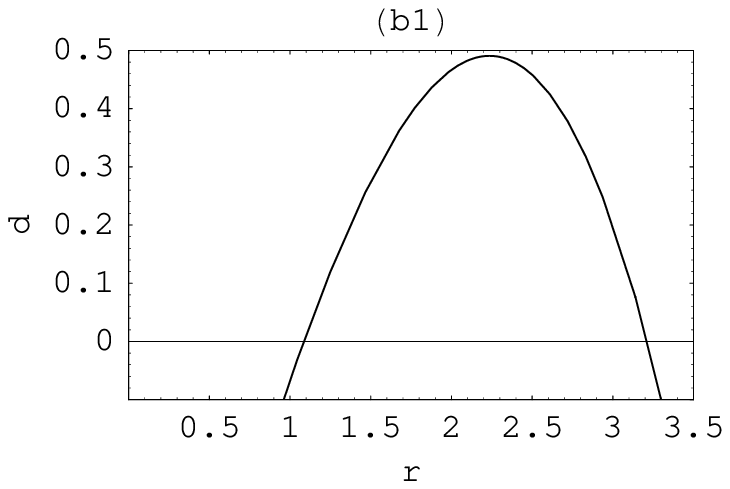}}\epsfysize=5cm
{ \epsfbox{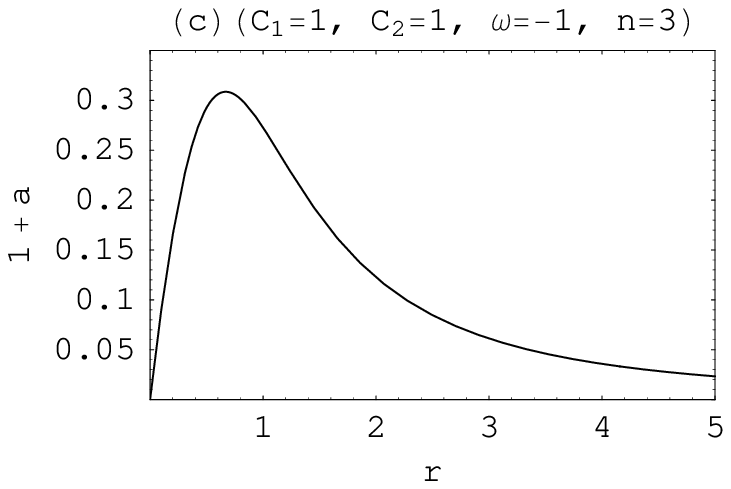}}\epsfysize=5cm { \epsfbox{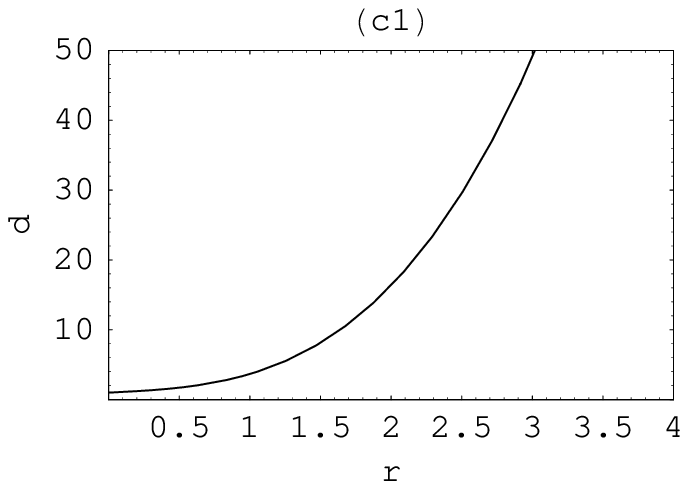}}\caption{The
behavior of $1+a(r)$ and $d(r)$ for different choices of
constants.} \label{i1}
\end{figure}

\begin{figure}\epsfysize=5cm
{ \epsfbox{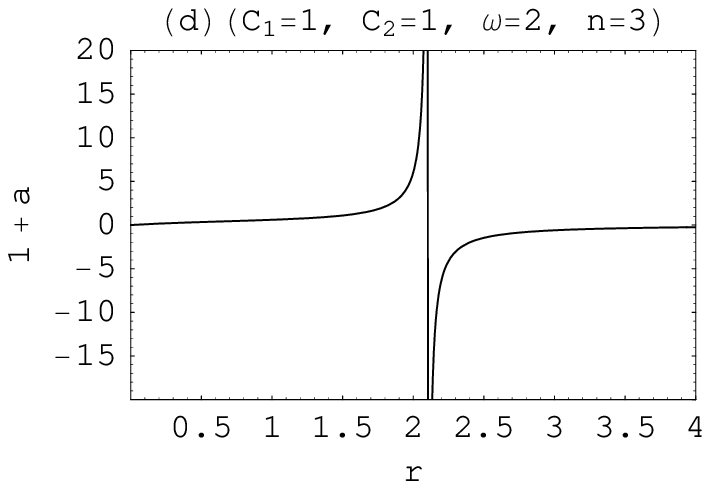}}\epsfysize=5cm { \epsfbox{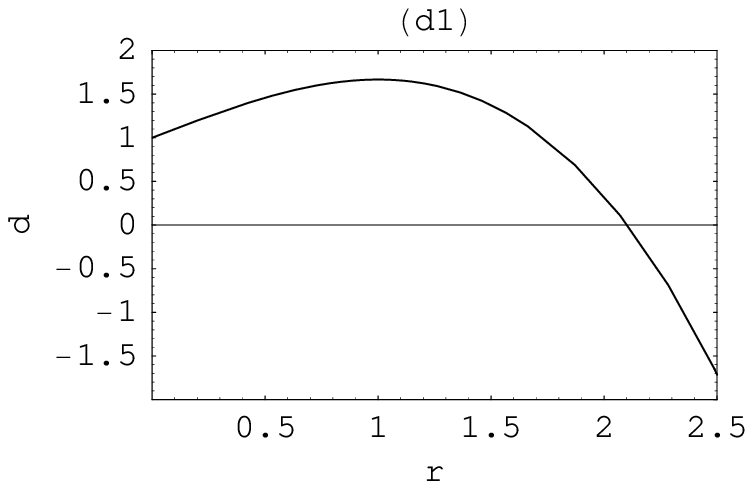}}\epsfysize=5cm
{ \epsfbox{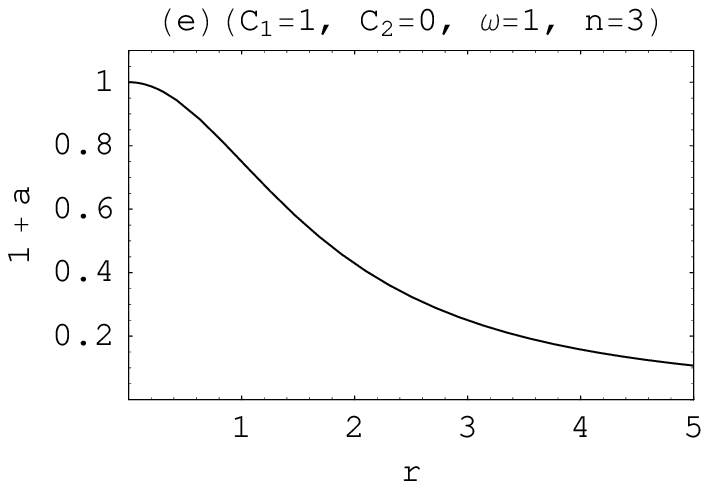}}\epsfysize=5cm { \epsfbox{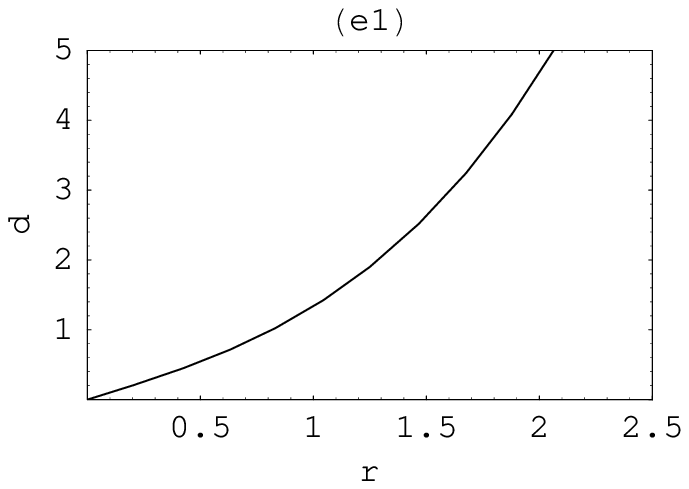}}\epsfysize=5cm
{ \epsfbox{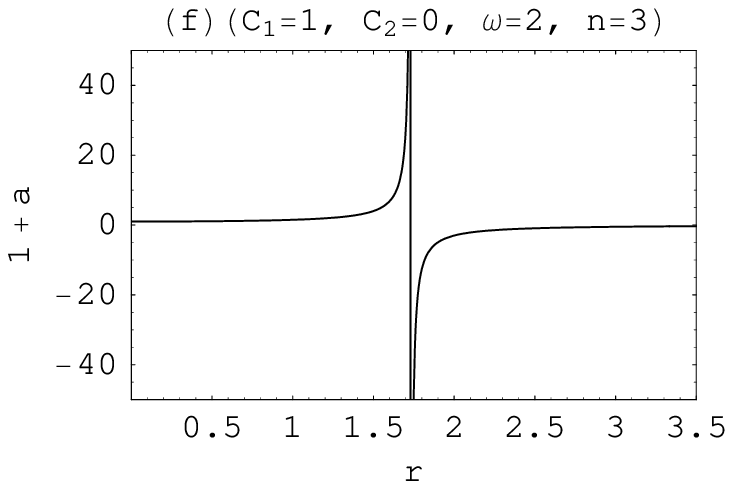}}\epsfysize=5cm { \epsfbox{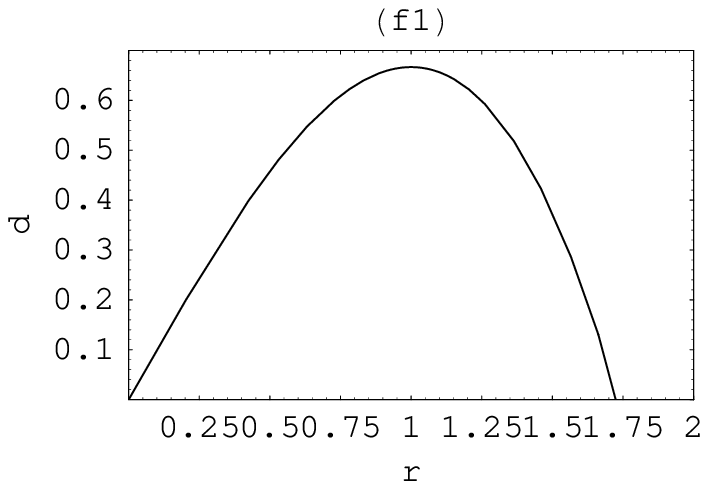}}\caption{The
behavior of $1+a(r)$ and $d(r)$ for different choices of
constants. } \label{i2}
\end{figure}

From equation (\ref{b2}) we have

\begin{equation}
\mathrm{d}(r) =0 \Rightarrow
C_{1}^2(1-\frac{n-1}{n}\omega)r^{n}+r^{n-2}+C_2=0,
 \label{db2}
\end{equation}
where $C_1$ and $C_2$ are in integration constants. Since we could
not solve this equation for general $n$ analytically, we plot
$d(r)$ against $r$ to show the different geometries this solution
could represent. We discuss a few specific solutions:

(a) For $ C_1=1$, $C_2=-1$, $\omega=-4$, we see from Fig.\ref{i1}
that we have a lower limit on $r$ which corresponds to the throat
radius of the wormhole and we see also that there is no upper
limit on $r$ which reminds us that we have an open spacetime.

(b) For $ C_1=1$, $C_2=-1$, $\omega=1.6$, we have lower and upper
limits on $r$. The lower limit corresponds to the throat radius of
the wormhole and the upper limit signifies a closed spacetime. In
this case, we have a wormhole in a closed universe.

(c) $ C_1=1$, $C_2=1$, $\omega=-1$, represents a naked singularity
in an open cosmological background, because the Kretschman scalar
blows up at the origin ($r=0$) and we see that the coordinate $r$
can reach the origin.

(d) For $ C_1=1$, $C_2=1$, $\omega=2$, the solution represents a
naked singularity again but this time in a closed universe.

(e) $ C_1=1$, $C_2=0$, $\omega=1$, leads to a maximally symmetric,
open spacetime which corresponds to the $FRW$ metric. For other
values of $\omega$, too, we have the same behavior.

(f) Finally, the choice $ C_1=1$, $C_2=0$, $\omega=2$, corresponds
to a closed, $FRW$ universe.

It is worth seeking the possibility of wormhole solutions with
$\omega>500$ which is motivated by  solar system observations. One
can see that with suitable choices of $C_1$, $C_2$ and
$\omega>500$ it is possible to have a wormhole geometry in a
closed spacetime, although with an exotic matter.

 An interesting point to note is that the value of the $\omega$
parameter affects the geometry of the spacetime and changing
$\omega$ from values smaller than ($\frac{n}{n-1}$) to larger
values will change the open geometry to the closed one, but this
change doesn't affect the expansion rate $R(t)$ or scalar field
$\phi(t)$. From (\ref{scalfac2}), one can see that the wormhole
solutions live in a spacetime which expands as $R(t)\propto t$.
This scale factor corresponds to the border line between
accelerating and decelerating universes. We can also look at the
limit $(\omega\rightarrow \infty)$ of this class of our solutions
and see that in this case, the limit does not approach the GR
solutions.

It is worth looking at the Ricci scalar for this class of
solutions. It turns out that the Ricci scalar is given by

\begin{equation}
{\cal R}=\frac{(n-1)^2\,\omega}{t^2}.
 \label{ricci2}
\end{equation}

It can be seen that the  Ricci scalar is spatially constant and
decreases with time.

\subsection{Wormhole's Two-Way Traversibility }

Perhaps the most interesting property of a wormhole is its
traversibility. Let us consider this problem in a short
discussion. The wormholes presented in this paper are traversable.
We present two reasons here. The first reasoning is based on the
redshift of a signal emitted at the comoving coordinate $r_{1}$
and received by a distant observer. Using the metric (\ref{metn})
and for a radial beam, we obtain

\begin{equation} \label{red1}
\frac{dt}{R(t)}=\left[1+a(r)\right]dr.
\end{equation}

Using this relation for two signals separated by $\tau_{0}$ in
time when emitted (and $\tau$ when detected), we obtain

\begin{equation} \label{red2}
\frac{\tau}{\tau_{0}}=1+z=\frac{R(t_{0})}{R(t_{1})},
\end{equation}
in which $R(t_{0})$ is the scale factor at the time of
observation, and $R(t_{1})$ is the scale factor at the time of
emission. This leads to exactly the same relation as the
cosmological redshift relation which shows that the wormhole does
not introduce extra (local) redshift. Light signals, therefore can
travel to the both sides of the throat and there is no horizon.

The second argument is based on the geodesic equation, which -for
the metric (\ref{metn})- leads to
\begin{equation} \label{red3}
\frac{d^{2}r}{d\lambda^{2}}+\frac{1}{2}\frac{a^{\prime}(r)}{1+a(r)}\left(\frac{dr}{d\lambda}\right)^2+2\frac{\dot
R}{R}\frac{dt}{d\lambda}\frac{dr}{d\lambda}=0
\end{equation}
and
\begin{equation} \label{red4}
\frac{d^{2}t}{d\lambda^{2}}+R\dot
R(1+a(r))\left(\frac{dr}{d\lambda}\right)^2=0,
\end{equation}
in which $\lambda$ is an affine parameter along the geodesic. The
first equation has the following first integral

\begin{equation} \label{red5}
\frac{dr}{d\lambda}=\frac{C}{R^2\sqrt{1+a(r)}}.
\end{equation}

Since the proper distance element is $\delta l=R\sqrt{1+a}\delta
r$, we see that there is no radial turning point and any particle
can move in either radial directions at any point near to the
wormhole, which clearly shows that the wormhole is traversable.

\section{Energy-Momentum Tensor and the weak energy condition\label{energy}}
Let us consider the energy-momentum tensor for different classes
of our solutions. Some points are interesting to mention about the
energy-momentum tensors which needed for the two cases. Since we
are looking for spherical structures in a cosmological background,
$P_{t}$, $P_{r}$ and $\rho$ should become almost $r$-independent
at large $r$. One can easily see that our different solutions have
this asymptotic behavior.

As we mentioned in the introduction, the main objection against
the plausibility of a wormhole solution is that the
energy-momentum tensor which supports this geometry violates the
weak energy condition $(WEC)$. Here we are interested in
investigating WEC for the presented classes of solutions.

The weak energy condition requires

\begin{equation}\label{E10}
T_{\mu \nu}u^{\mu}u^{\nu} \geq0
\end{equation}
for every nonspacelike $u^{\mu}$ which leads to \cite{22}

\begin{equation}\label{E11}
\rho \geq0      , \  \rho+P_{r} \geq0,\ \and \rho+P_{t}\geq0 .
\end{equation}

 In order to investigate the WEC for class (I), we don't need to consider all the above relations.
 One can easily see that for class (I), according to (\ref{rho1}), the first equation ($\rho \geq 0$) is not satisfied. Then the
 wormhole solution presented in this case definitely violates the
 WEC.

 Using (\ref{E11}) and the relations (\ref{rho2}-\ref{pt2}) for
 the second class of solutions we have

\begin{figure}\epsfysize=5cm
{ \epsfbox{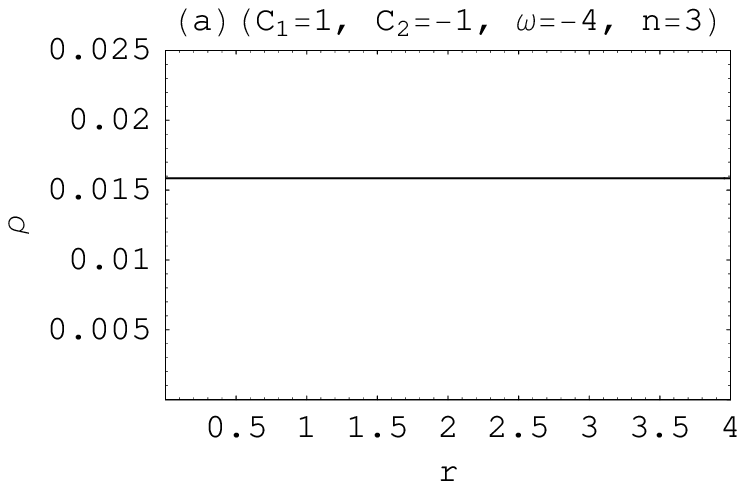}}\epsfysize=5cm {
\epsfbox{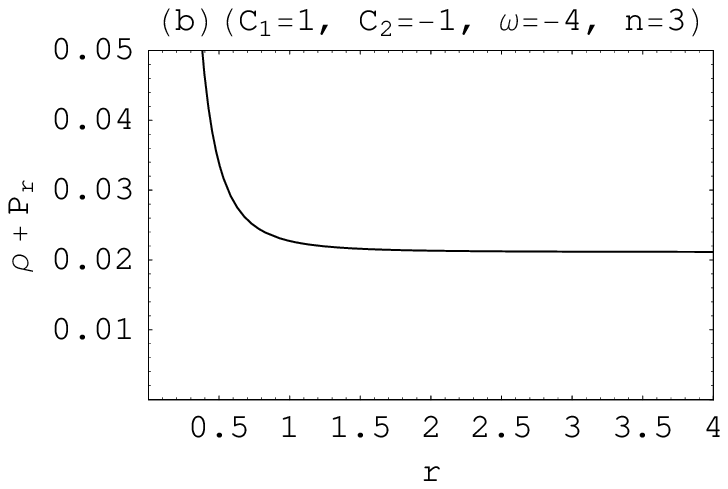}}\epsfysize=5cm { \epsfbox{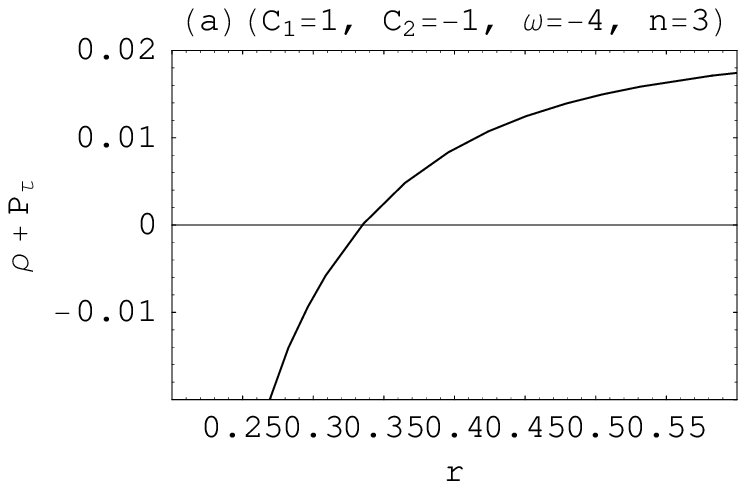}}\epsfysize=5cm
{ \epsfbox{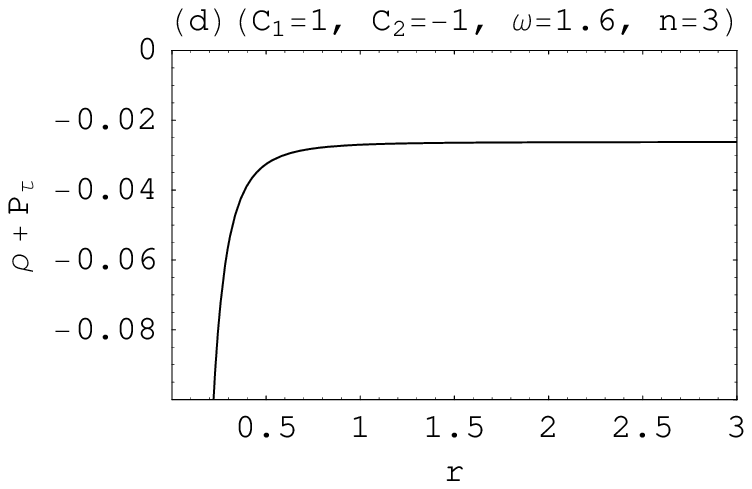}}\epsfysize=5cm \caption{The energy condition
factors are plotted against $r$. The first three plots are for
wormhole in an open universe. The last figure (d) corresponds to
$\rho+P_t$ for a wormhole in a closed universe.} \label{i3}
\end{figure}

 \begin{equation}\label{E12}
    \rho \geq 0 \Rightarrow -\frac{(n-1)(n+(n-1)\omega)}{64\,\pi^2\,
G}At^{-(n+1)} \geq 0,
\end{equation}

\begin{equation}\label{E13}
 \rho + P_{r} \geq 0 \Rightarrow  -\frac{(n-1)(2C^2_{1}(n+1)(n+(n-1)\omega)+n(n-2)C_2r^{-n})}{128\,\pi^2\,G\,n\,C_{1}^2}\,At^{-(n+1)} \geq 0,
\end{equation}
and
\begin{equation}\label{E14}
\rho + P_{t} \geq 0  \Rightarrow
\frac{(-2C^2_{1}(n-1)(n+1)(n+(n-1)\omega)+n(n-2)C_2r^{-n})}{128\,\pi^2\,G\,n\,C_{1}^2}\,At^{-(n+1)}
\geq 0.
\end{equation}

One can see from Fig.(\ref{i3}) that the weak energy condition can
be satisfied in the case of wormhole in an open universe which
corresponds to $\omega<0$, and for suitable values of $\omega$.
Here, one should note that the supporting matter energy-momentum
tensor satisfies the WEC but the effective energy-momentum tensor
which includes the BD scalar field and the matter energy-momentum
tensor does violate the WEC. We can see from Fig.(\ref{i3}) part
(d) that $\rho+P_t$ is everywhere negative and the wormhole in a
closed universe always violates the WEC. It is also notable to
mention that the wormhole solutions with $\omega>500$ (as favored
by observation) do violate the WEC.

\section{  Summary And Conclusion}

In this paper, we studied exact (n+1)-dimensional, spherical
geometries in a cosmological background, in the framework of the
BD theory. To this goal, we used a metric which is a simple
extension of RW metric and considered a source with traceless
energy-momentum tensor. Using these ansatzen, two classes of
solutions were found. Class (I), represented wormhole geometries
in an open universe. For this class of solutions, the scale
factor, BD scalar field and the energy-momentum tensor needed to
support the geometry were calculated. It turned out that, changing
$\omega$ affected the expansion rate $R(t)$ and the scalar field
$\phi(t)$. We also calculated the Ricci scalar and saw that it was
spatially constant and decreased with time. The second class of
solutions was richer than the first one. We classified the
solutions in different categories with distinct geometries. Two
classes of solutions were shown to represent Lorentzian wormholes
in open and closed universes. Two spacetimes containing naked
singularity and two maximally symmetric spacetimes were found. For
the second class of solutions, we found that the scale factor was
proportional to time which corresponded to the border line between
decelerating and accelerating universes. In this class, -despite
the previous class- we saw that changing $\omega$ affected the
global geometry of spacetime and did not change the expansion rate
$R(t)$ and scalar field $\phi(t)$. Once again, for the second
class, we showed that the Ricci scalar was spatially constant and
decreased with time. For both classes of solutions we saw that
$\rho$ was only time dependent while $P_r$ and $P_t$ depended on
both $r$ and $t$. In general, one expects that since we are
dealing with an inhomogeneous cosmology, pressure and density
should depend on both space and time coordinates. Although this is
the case for the pressure, it was seen that the density depended
only on $t$ as in the homogeneous cosmological models. This
unexpected result which came from the field equations, seems to be
a property of the BD field equations and does not occur in the
Einstein gravity. In addition to that, if we define an average
pressure according to ($\bar{P}=\frac{P_r+(n-1)P_t}{n}$), we see
that $\bar{P}$ depends only on time, too. We also investigated the
traversibility of the wormhole solutions and showed that the
presented wormholes were traversable. Finally, we considered the
energy-momentum tensor for different classes of solutions and
investigated the WEC for wormhole solutions. The solutions led to
energy-momentum tensors which became almost $r$-independent as we
got far from the central object. We saw that the wormhole
solutions presented in class (I) violated the WEC. For the second
class, we found out that the matter energy-momentum tensor
supporting the wormhole solutions in an open universe with
suitable choice of constants could completely satisfy the WEC
everywhere. However, the effective energy-momentum tensor was seen
to violate the WEC even in this case. Wormhole living in a closed
universe always violated WEC.

 \acknowledgments{N.Riazi acknowledges the support of Shiraz University Research Council.}


\begin{thebibliography}{99}

\bibitem{jordan} P. Jordan, Schwerkraft und Weltall, Vieweg (Braunschweig) 1955.

\bibitem{bd} C. H. Brans , R. H. Dicke, Phys. Rev. \textbf{124}, 925 (1961).

\bibitem{reff} L. Flamm, Physik. Zeitschr. 17, 448 (1916).

\bibitem{ERB}  A. Einstein, N. Rosen, Phys. Rev. \textbf{48}, 73 (1935).

\bibitem{wheeler}  J. A. Wheeler, Ann. Phys. \textbf{2}, 604 (1957).

\bibitem{MT}  M. S. Morris, K. S. Thorne , Am. J. Phys. \textbf{56} (1988)
395.

\bibitem{MTY}  M. S. Morris, K. S. Thorne and U. Yurtsever, Phys. Rev. Lett. \textbf{61}, 1446 (1988).

\bibitem{bord}  A. Borde, Class. Gravitation \textbf{4}, 343 (1987).

\bibitem{haw}  S. W. Hawking, Phys. Rev \textbf{D 46}, 603 (1992).

\bibitem{visser} M. Visser, Lorentzian wormholes: form Einstein to
Hawking, I. E. P. Press, Woodsbury, N. Y. 1995.

\bibitem{11}  H. Epstein, V. Glaser, A. Jaffe, Nuvo Cimento \textbf{36}, 1016(1965).

\bibitem{14}  N. Riazi, J.Korean Astron. Soc.\textbf{29}, S283(1996).

\bibitem{15}  N. Riazi, Astrophys. Space Sci. \textbf{283}, 231 (2003).

\bibitem{16}  J.M. Maldacena and L. Maoz, JHEP 0402, 053
(2004). hep-th/0401024.

\bibitem{17} E. Bergshoeff, A. Collinucci, A. Ploegh, S. Vandoren and T.
Van Riet, JHEP 0601, 061 (2006). hep-th/0510048.

\bibitem{18} N. Arkani-Hamed, J. Orgera and J. Polchinski, JHEP 0712, 018
(2007). hep-th/0705.2768.

\bibitem{19} A. Bergman and J. Distler, hep-th/0707.3168.

\bibitem{20} E. Bergshoeff, W. Chemissany, A.
Ploegh, M. Trigiante and T. Van Riet, hep-th/0806.2310.

\bibitem{accelreff} S. Perlmutter et al., Astrophys. J. 517 (1999) 565-586,
astro-ph/9812133; A. G. Riess et al., Astron. J. 116 (1998)
1009-1038, astro-ph/9805201; Astrophys.J. 560 (2001) 49-71,
astro-ph/0104455.

\bibitem{agnes} A. G. Agnese and M. La Camera, Phys. Rev. \textbf{D 51}, 2011
(1995).

\bibitem{nandi} K. K. Nandi, A. Islam and J. Evans, Phys. Rev. \textbf{D 55}, 2497
(1997).

\bibitem{mat} T. Matos and D. N´u˜nez, gr-qc/0508117 and references therein; T.
Matos, Gen. Rel. Grav. 19, 481 (1987).

\bibitem{kim} F. He and S-W. Kim, Phys. Rev. \textbf{D 65}, 084022
(2002).

\bibitem{cramer} J. G. Cramer, et. al., Phys. Rev. \textbf{D 51}, 3117(1995).

\bibitem{harko} T. Harko, Z. Kovacs, F. S. N. Lobo,  Phys. Rev.
\textbf{D 78}, 084005(2008).

\bibitem{guth}  A. H. Guth, Phys. Rev. \textbf{D 23}, 347 (1981).

\bibitem{riazinasr}  N. Riazi, B. Nasr , Astrophys. Space Sci. \textbf{271}, 237(2000).

\bibitem{22}  S. Carroll, Spacetime and Geometry: An Introduction to General Relativity,
Addison Wesley, USA (2004).

\end{thebibliography}
\end{document}